\documentclass[apjl]{emulateapj}
\begin{document}
\title{Millimeter Flares and VLBI Visibilities from Relativistic Simulations of Magnetized Accretion onto the Galactic Center Black Hole}
\shorttitle{GRMHD Study of Sgr A*}
\shortauthors{Dexter, Agol \& Fragile}
\author{Jason Dexter}
\affil{Department of Physics, University of Washington, Seattle, WA 98195-1560, USA}
\email{jdexter@u.washington.edu}
\author{Eric Agol}
\affil{Department of Astronomy, University of Washington, Box 351580, Seattle, WA 98195, USA}
\author{P. Chris Fragile}
\affil{Department of Physics \& Astronomy, College of Charleston, Charleston, SC 29424}
\keywords{accretion, accretion disks --- black hole physics --- radiative transfer --- relativity --- galaxy: center}
\slugcomment{ApJL, accepted}
\begin{abstract}
The recent VLBI observation of the Galactic center black hole candidate Sgr A* at 1.3mm shows source structure on event-horizon scales. This detection enables a direct comparison
of the emission region with models of the accretion flow onto the black hole. We present the first results from time-dependent radiative transfer of general relativistic MHD simulation data, and compare simulated synchrotron images at black hole spin $a=0.9$ with the VLBI measurements. After tuning the accretion rate to match the millimeter flux, we find excellent agreement between predicted and observed visibilities, even when viewed face-on ($i \lesssim 30^\circ$). VLBI measurements on $2000$--$3000$km baselines should constrain the inclination. The data constrain the accretion rate to be ($1.0$--$2.3$)$\times10^{-9} M_{\sun} \mathrm{yr}^{-1}$ with $99\%$ confidence, consistent with but independent of prior estimates derived from spectroscopic and polarimetric measurements. Finally, we compute light curves, which show that magnetic turbulence can directly produce flaring events with $.5$ hour rise times, $2$--$3.5$ hour durations and $40$--$50\%$ flux modulation, in agreement with observations of Sgr A* at millimeter wavelengths.
\end{abstract}
\maketitle

\begin{figure*}
\epsscale{1.2}
\plotone{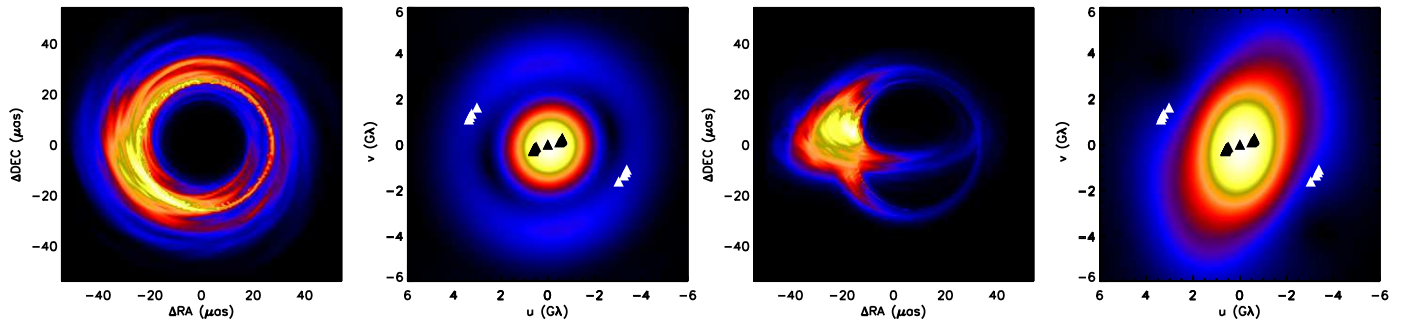}
\caption{\label{imgs}Best fit ray traced images and corresponding 2D, scatter broadened and rotated visibilities for $i=10^\circ$ (left) and $i=70^\circ$ (right). The intensities are scaled linearly to the maximum of each panel (from blue to red to yellow), and the UV plane locations of the VLBI observations from \citet{doeleman2008} are overplotted on the visibilities as triangles. The ring of peak brightness at a radius of $\sim25 \mu$as in the $i=10^\circ$ image is due to the transition between bound and unbound photons. Despite the complex structure of the $i=70^\circ$ image, the visibility is similar in shape to that in \citet{broderick2009} Fig. 2.}
\end{figure*}

\section{Introduction}

Due to its large angular size, the Galactic supermassive black hole candidate Sgr A* is a promising laboratory for precision black hole astrophysics using very long baseline interferometry (VLBI) at millimeter wavelengths. Previous measurements \citep{shen05,doeleman2001,krich98} at 7 and 3.5mm found a small instrinsic source size for Sgr A* (100--200 $\mu$as FWHM), but at those wavelengths interstellar scattering is the dominant contribution to the observed size. Recent measurements by \citet{doeleman2008} at 1.3mm are the first at short enough wavelengths to avoid contamination due to interstellar scattering and at long enough baselines to achieve event-horizon scale resolution. These measurements have been used by \citet{broderick2009} to constrain the black hole and accretion flow parameters in terms of a 1.5D, nonrelativistic, time-steady radiatively inefficient accretion flow model (RIAF; \citealt{yuanquataert2003}). 

Sgr A* is also known to exhibit multiwavelength, polarized flares (e.g., \citealt{zadeh2007,marrone2008,eckart2008pol,doddseden2009}). This activity has been modeled as hotspots orbiting in the inner radii \citep{broderickloeb2005}, adiabatically expanding blobs \citep{yusefzadeh2006,eckart2008sim} and emission from a short, mildly relativistic jet \citep{falcke1993}. 

As a complementary approach to the relatively simple models outlined above, we perform relativistic radiative transfer on data from a recent 3D, general relativistic MHD (GRMHD) simulation of a black hole accretion disk \citep{fragile2007}, and compare simulated synchrotron images to the VLBI observations. In addition, we produce light curves at 1.3mm as a first step towards explaining Sgr A*'s flaring activity in terms of magnetic turbulence. 

In contrast to the results from RIAF modeling, we find that for the single black hole spin considered ($a=0.9$), \emph{all} inclinations provide excellent fits to the VLBI data. Visibility profiles from low (face-on) inclinations ($i \lesssim 30^\circ$) resemble ``ring" models (e.g., \citealt{doeleman2008}), whereas those at higher (edge-on) inclinations are more Gaussian. Ongoing work with increased array sensitivity should be able to distinguish between these two types of visibility profiles.

Detection of a ring-like profile would provide an important constraint not only on the geometry of Sgr A*, but also on the necessary complexity of accurate accretion flow models. These results also demonstrate the power of mm VLBI to distinguish between accretion models, as well as the model-dependence of attempts to extract black hole parameters.

\section{Methods}

\subsection{Simulation Data}

\citet{fragile2007} presented global GRMHD simulations of black hole accretion disks. They used a spherical grid with polar Kerr-Schild coordinates, and the main untilted simulation had an effective resolution of $128^3$, except near the poles which are underresolved. The torus was initialized with the analytic, time-steady, axisymmetric solution of \citet{devilliers2003} and threaded with a weak poloidal magnetic field with minimum $P_{gas}/P_{mag}=10$ initially. The magnetorotational instability (MRI) arose naturally from the initial conditions, and the disk quickly became fully turbulent. The simulation was evolved for $\sim8000$M, or $\sim$40 orbits at $r=10$M in units with $G=c=1$. Only data from the final $2/3$ of the simulation are used in this letter, once the disk is fully turbulent and transient effects from the initial conditions have died down.

This simulation evolved an internal energy equation, and injected entropy at shocks. Such a formulation does not conserve energy, and produces a more slender, cooler torus than conservative simulations which capture the heat from numerical dissipation of magnetic fields \citep{fragile2009}.

\subsection{Ray Tracing}

We performed relativistic radiative transfer on the simulation data via ray tracing. Starting from an observer's camera, rays are traced backwards in time assuming they are null geodesics (geometric optics approximation), using the public code described in \citet{dexteragol2009}. In the region where rays intersect the accretion flow, the radiative transfer equation is solved along the geodesic \citep{broderick2006} in the form given in \citet{fuerstwu2004}, which then represents a pixel of the image. This procedure is repeated for many rays to produce an image, and at many time steps of the simulation to produce time-dependent images (movies). Light curves are computed by integrating over the individual images.

To calculate fluid properties at each point on a ray, the spacetime coordinates of the geodesic are transformed from Boyer-Lindquist to the Kerr-Schild coordinates used in the simulation. Since the accretion flow is dynamic, light travel time delays along the geodesic are taken into account. Data from the sixteen nearest zone centers (eight on the simulation grid over two time steps) were interpolated to each point on the geodesic. Between levels of resolution near the poles, data from the higher resolution layer were averaged to create synthetic lower resolution points, which were then interpolated. Little emission originates in the underresolved regions of the simulation.

In this work, we consider synchrotron emission from Sgr A* using the formulae from \citet{maha} and neglecting polarization. The simulation provided mass density, pressure, velocity and magnetic field in code units. These were converted into cgs units following the procedure described in \citet{schnittman2006}, except that we scaled the torus mass to match the observed flux of Sgr A* at 1.3mm. The length- and time-scales were set by the black hole mass, taken to be $4\times10^6 M_{\sun}$, at a distance of $8$ kpc.

The temperature was taken to be thermal, with the ion and electron temperatures equal. Many authors have used two temperature models of Sgr A*, with weak coupling between the dynamically important ions and the cooler electrons. However, our single temperature model fits the observed flux from Sgr A* at reasonable accretion rates $\sim 10^{-9} M_{\sun} \mathrm{yr}^{-1}$ and produces a spectrum near millimeter wavelengths that is consistent with observations. In this model, the gas is optically thin except at high inclinations during flares. The gas temperature is $\sim10^{11} K$ in the regions of peak synchrotron emissivity, which is consistent with peak electron temperatures in RIAF models. We also followed \citet{goldston2005} and produced images with $(T_e+T_i)/T_e=10$, keeping the total gas pressure fixed. These models became optically thick before reaching the observed 1.3mm flux, and were inconsistent with the location of the peak in the quiescent spectrum of Sgr A*.

\begin{figure}
\epsscale{1.1}
\plotone{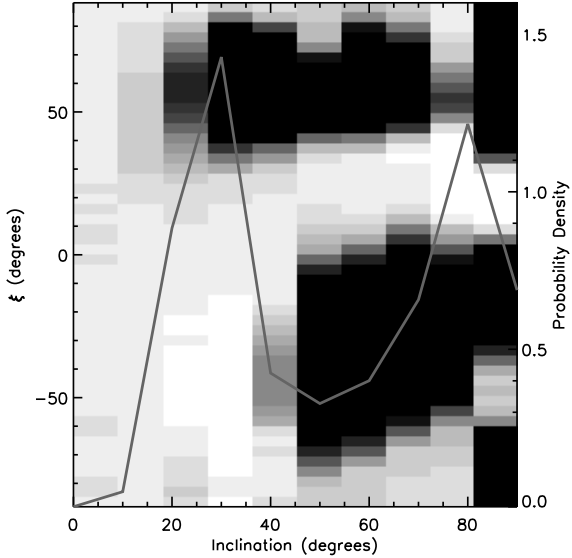}
\caption{\label{chi2}Grid of best fit reduced effective $\chi^2$ values vs. inclination and sky orientation $\xi$. The scale is from $\chi^2=1$ (white) to $2.5$ and greater (black). We find excellent fits at low inclinations, which are roughly independent of sky orientation. At high inclinations, our results show a similar shape to that found by \citet{broderick2009}. Overplotted is the probability density vs. inclination, marginalized over observer time, accretion rate and sky orientation.}
\end{figure}

\begin{figure}
\epsscale{1.1}
\plotone{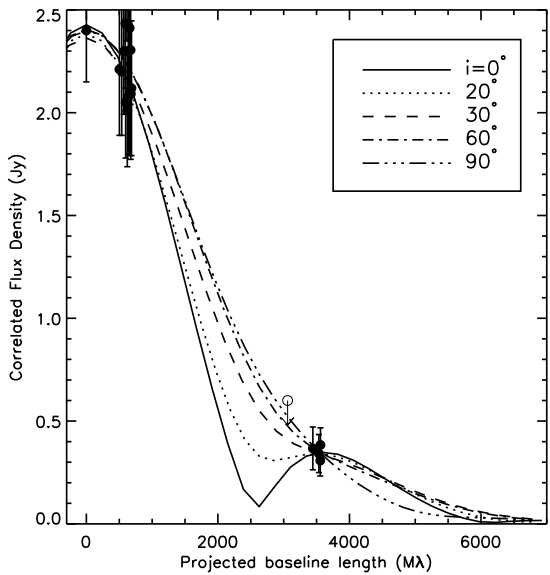}
\caption{\label{profs}Best fit visibility profiles for low and high inclinations, plotted along the line in the UV plane corresponding to the average location of the long baseline measurements from \citet{doeleman2008}. The visibilities naturally divide into two types. At low inclinations, the nearly circular shadow leads to a minimum in the visibility profile, similar to the ring model in \citet{doeleman2008}. At inclinations $\gtrsim 30^\circ$, the profiles monotonically decrease with baseline length. A detection in place of the current upper limit (open circle with arrow) should favor one set of profiles.}
\end{figure}

\begin{figure}
\epsscale{1.1}
\plotone{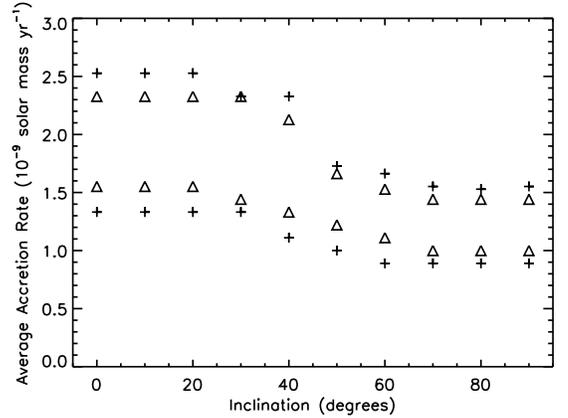}
\caption{\label{mdot}Upper and lower limits to the average accretion rate as a function of inclination at $68\%$ (triangles) and $99\%$ (plus signs) confidence from probability distributions marginalized over observer time and sky orientation for the $a=0.9$ simulation. The narrow allowed regions demonstrate the ability of VLBI observations in conjunction with GRMHD models to constrain the accretion rate of Sgr A*. Other simulations may produce different ranges of allowed accretion rates.}
\end{figure}

\section{Results}

\subsection{Visibility Fitting}

Our fitting procedure closely mirrors that described in \citet{broderick2009}. Simulated images were produced for a range of inclination angles, time-averaged accretion rates, and observer times (sets of simulation time steps). The images were then averaged over 10 minute observation periods (8 simulation time steps) to match the integration times used in \citet{doeleman2008}, padded sufficiently to resolve the shortest baselines, and Fourier transformed to 2D visibilities. The visibilities were rotated to a range of different orientations, $\xi$, the projection of the black hole spin axis on the sky measured E of N. Interstellar scattering was then applied by multiplication by an elliptical Gaussian as described in \citet{fish2009} using the fits from \citet{bower2006}. Sample images and scatter-broadened visibilities are shown in Fig. \ref{imgs}. At low inclinations, the emission region is essentially a ring around the shadow, which causes a minimum in the 2D visibilities. For higher inclinations $\gtrsim 30^\circ$, Doppler beaming causes the emission to be concentrated on one side of the black hole. The visibilities are then approximately elliptical, with the minor (major) axis corresponding to the major (minor) axis of the emission region.

Visibilities were interpolated to the detections in \citet{doeleman2008}, and an effective $\chi^2$ was computed as defined in \citet{broderick2009}. Grids of minimum effective, reduced $\chi^2$ are shown in Fig. \ref{chi2} over inclination angle and sky orientation. At inclinations $i \geq 30^\circ$, our results are in at least qualitative agreement with \citet{broderick2009}. The orientation is well constrained at high inclinations, since the general image shape is nearly static and asymmetric, and the long baseline VLBI measurements pick out a specific orientation for the visibility ellipse. The two distinct bands of good fits are due to the (approximate) up-down symmetry of the image. However, RIAF models ruled out inclinations $i \lesssim 30^\circ$, whereas for our GRMHD models, good fits (reduced $\chi^2 \lesssim 1.2$) are possible at all inclinations. This is especially evident from the curve of $p$($i$) vs. $i$ overplotted in Fig. \ref{chi2}. Inclinations $i \lesssim 20^\circ$ are less probable due to the $\sin{i}$ prior and large variation of probability density with observer time. Although the best fit $\chi^2$ is roughly the same for low inclinations at all $\xi$, at any given observer time good fits are restricted to less than half the range of sky orientations.

Visibility profiles from the best fits at many inclinations are shown in Fig. \ref{profs}. The profiles represent the 2D visibility plotted along the line representing the average angle in the UV plane of the long baseline measurements from \citet{doeleman2008}. Also plotted are the data from that work. At low inclinations, the profile reaches a minimum as described above, whereas at $i \geq 30^\circ$, the profile decreases monotonically with baseline length. Our best fit face-on visibility profile is almost identical to that of the ring model from \citet{doeleman2008}. 

Our images are strongly peaked at the circular photon orbit, as can be seen in Fig. \ref{imgs}. Low inclination ($i \lesssim 30^\circ$) images produced from a single time step of the energy conserving simulation by \citet{mckinneyblandford2009} are similar to that in Fig. \ref{imgs}, suggesting that our results are not due to the nonconservative nature of the simulation. It is not surprising that the VLBI observations are well fit by face-on disks. In addition to the ring model, a delta function intensity profile at the circular photon orbit roughly matches the observations, as does the spherical accretion model from \citet{dexteragol2009} scaled to the mass of Sgr A*. In fact, any model sharply peaked at the circular photon orbit should approximately match the visibility data. This is because the ratio of the zeroth to first maximum in the visibility for such a model is roughly the same as that of the zeroth order Bessel function, and agrees with the ratio of the visibility amplitude at $\sim3500 \mathrm{M}\lambda$ to the total flux.

\begin{figure}
\epsscale{1.2}
\plotone{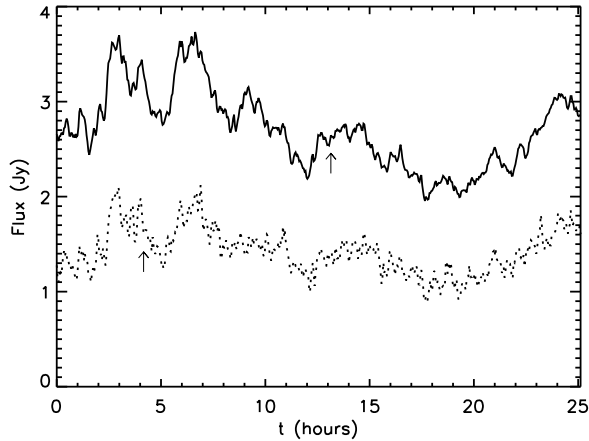}
\caption{\label{lcurve}Light curves of the last $2/3$ of the untilted GRMHD simulation from \citet{fragile2007} at inclinations of $10^\circ$ (top) and $70^\circ$ (bottom). The $i=70^\circ$ light curve has been shifted downwards by $1.5$ Jy for clarity. Both light curves exhibit consecutive flares starting at $t \sim 2$ hours, which are consistent with those observed from Sgr A* at mm wavelengths. They are more prominent at $i=10^\circ$ due to higher optical depth between the observer and the flaring region at high inclinations. The arrows denote the 10 minute intervals corresponding to the best fit images shown in Fig. \ref{imgs}.}
\end{figure}

If this GRMHD simulation approximates the correct model for Sgr A*, we can also constrain the accretion rate for the single spin value used. Upper and lower limits for $68\%$ and $99\%$ confidence from probability distributions marginalized over observer time and sky orientation are shown in Fig. \ref{mdot}. Doppler beaming leads to larger fluxes at high inclinations, and hence a lower accretion rate is necessary to match the observed flux. The limits are also more stringent at high inclinations, where the flux is less variable (see Fig. \ref{lcurve}). Further marginalizing over inclination, we constrain the accretion rate to be ($1.0$--$2.3$)$\times10^{-9}M_{\sun} \mathrm{yr}^{-1}$ at $99\%$ confidence.  This constraint on the accretion rate of Sgr A*, while model-dependent, only uses the VLBI visibility measurements.

\subsection{Variability}

A major advantage of using GRMHD simulations in place of RIAF models is the ability to probe variability from the same dynamical model used to produce spectra and visibilities. Previous work has used hotspots \citep{broderickloeb2005,doeleman2009} to study variability, but this introduces extra free parameters, complicating attempts to extract black hole parameters. The light curves of our Sgr A* models at 1.3mm shown in Fig. \ref{lcurve} are the first from simulation data to consider synchrotron emission, and the first to include the time delay along geodesics.

The variability on short timescales is slightly more noticeable for the $i=70^\circ$ case. Interestingly, though, the longer time scale flaring behavior is more pronounced in the $i=10^\circ$ light curve. The power spectra are well described by red noise spectra with power law index $\Gamma=2.4$, $1.7$ for $i=10^\circ$, $70^\circ$. These are both steeper than the observed power law index of $\Gamma=1$ in \citet{mauerhan2005} at 3mm wavelength. 

The twin flares occur simultaneously in both light curves, rise over half an hour and last $2$--$3.5$ hours. The flux modulation is $50\%$ ($40\%$) at $i=10^\circ$ ($70^\circ$), measured from the peak of the flare to the average of the light curve immediately preceding it. All of these features are consistent with mm flare observations of Sgr A* \citep{eckart2008sim,marrone2008,zhao2003,li2009}. Since the flares are seen at low inclinations, they are not caused by Doppler shifts from hotspots. The peak intensity is attenuated at large inclination due to rising optical depth. The flares are caused by a rise in magnetic field strength near the midplane in the inner radii ($r \sim 2$--$6$M). They are not due to heat from magnetic dissipation since this is not possible in the simulation used here. Thus heating from magnetic reconnection is not necessary to produce the mm flares of Sgr A*.
 
\section{Discussion}

In this letter, simulated images of a single temperature synchrotron emission model from 3D, GRMHD simulations by \citet{fragile2007} have been fit to recent 1.3mm VLBI observations of Sgr A*. For the single black hole spin considered, the high inclination results are in at least qualitative agreement with those presented for 1.5D, stationary, non-relativistic models used by \citet{broderick2009}. However, we find that in contrast to previous work, images at low inclination provide excellent fits to the data, which are relatively insensitive to sky orientation, and present a qualitatively different visibility profile due to approximate circular symmetry. 

Radial intensity profiles from face-on images of recent RIAF models \citep{yuan2009} appear more extended than those from our GRMHD simulations. The minimum of the visibility profile for these images occurs at too short a baseline to match the VLBI data. The model in \citet{broderick2009}, on the other hand, has increasing number density up to the event horizon, in stark contrast to the results from the simulations, where the number density peaks at $r \sim 5M$, well outside of the event horizon at $r=1.4M$ and the marginally stable orbit at $r=2.3M$. Large number densities in the inner radii of the model from \citet{broderick2009} cause the ``shadow" to be much smaller than expected \citep{bardeen1973,falcke}, weakening the power on small scales (large baselines), and preventing a good fit to the VLBI data at low inclinations. 

Increased 1.3mm array sensitivity should favor either the low or high inclination profiles in Fig. \ref{profs}. Our predicted profiles would also be sensitive to measurements at 1.3mm along baselines of length $2000$--$3000$km. A future detection favoring a ring-like profile would constitute the first measurement of a black hole shadow.

The visibility analysis constrains the average accretion rate of the $a=0.9$ simulation to be ($1.0$--$2.3$)$\times10^{-9} M_{\sun} \mathrm{yr}^{-1}$ with $99\%$ confidence. While this constraint is model-dependent, it is independent of and consistent with constraints from spectral fitting \citep{yuanquataert2003} and polarization measurements \citep{agol2000,quataert2000,macquart2006,marrone2006,sharma2007}, and in good agreement with the previous spectral fits of single temperature GRMHD synchrotron images by \citet{noble2007}.

We have also produced the first light curves of synchrotron emission from GRMHD modeling of Sgr A*. At 1.3mm, flares arise naturally from magnetic turbulence, and are consistent with observations. Unfortunately our analysis is poorly suited for studying variability in other wavebands such as the radio, NIR or X-ray. In the radio, most the emission originates outside of the initial torus pressure maximum at $r=25M$, while the NIR and X-ray luminosity is produced by some combination of synchrotron emission from non-thermal electrons and Compton scattering of synchrotron photons. Modeling of both is left for future work. In addition, polarization measurements from flares are an important observational clue. Adding this capability to our relativistic radiative transfer code is in progress.

There are many uncertainties in the analysis presented in this letter. While GRMHD allows for a self-consistent solution of the primitive variables, it is unknown whether the MHD approximation is valid for the low densities at the Galactic center. In addition, there are a range of possible solutions, depending on the energy prescription used. Current simulations also neglect the interplay between ions and electrons, which should be handled self-consistently.

If Sgr A* is radiatively inefficient, simulations which conserve total energy are probably more appropriate. We estimate the time-averaged energy dissipation rate from numerical losses in the simulation as $\sim10^{37}$ ergs $\mathrm{s}^{-1}$ for an accretion rate of $1.6\times10^{-9} M_\sun \mathrm{yr}^{-1}$. This is larger than the bolometric luminosity of Sgr A*, and gives an effective radiative efficiency of $\sim.1$. The accretion flow is advection dominated, in that an order of magnitude more energy is lost to the black hole than to numerical reconnection. Face on synchrotron images from a single timestep of the 3D, conservative code in \citet{mckinneyblandford2009} were similar to those found here, suggesting that our results may not be specific to nonconservative simulations. 

Despite the uncertainties, relativistic radiative transfer is an important tool for connecting state of the art, 3D, global GRMHD simulations with observations. Our analysis shows that ongoing millimeter VLBI observations will not only constrain black hole parameters, but accretion models as well. Expanding this analysis to the full range of spins will help determine the model dependence of such future constraints. If Sgr A* is at a low inclination, mm VLBI may soon provide the first direct evidence of a black hole shadow.

In addition to the visibility fits and light curves considered here, the variability of Sgr A* can be probed through measurements of closure phases along triangles of baselines (e.g., \citealt{doeleman2009}). Identifying salient features of these closure phases is a goal of future work. If the flares seen here are robust features of mm synchrotron light curves of GRMHD simulations, multiwavelength and polarization measurements can be used to constrain emission mechanisms and the non-thermal electron distribution.

\begin{acknowledgements}
J.D. thanks Shep Doeleman for the VLBI data, and Omer Blaes for useful discussions. This work was partially supported by NASA grants 05-ATP05-96 and NNX08AX59H, and a graduate fellowship at the Kavli Institute of Theoretical Physics at the University of California, Santa Barbara under NSF grant PHY05-51164.
\end{acknowledgements}

\bibliographystyle{/Users/Jason/latex/apj}

\end{document}